\documentclass[12pt]{article}
\usepackage{amsfonts,graphicx,amssymb,amsmath,amsthm,epsfig}
\usepackage{float}
\allowdisplaybreaks
\begin{document}
\title{\bf Impact of Charge on the Stability of Pulsar SAX J1748.9-2021 in
Modified Symmetric Teleparallel Gravity}
\author{M. Sharif \thanks {msharif.math@pu.edu.pk}~and
\ Madiha Ajmal \thanks {madihaajmal222@gmail.com} \\
Department of Mathematics and Statistics, The University of Lahore,\\
1-KM Defence Road Lahore-54000, Pakistan.}

\date{}
\maketitle
\begin{abstract}
In this paper, we explore the effect of charge on the stability of
pulsar star SAX J1748.9-2021 in $f(Q)$ gravity, where $Q$ represents
non-metricity. For this purpose, we apply the Krori-Barua metric
ansatz with anisotropic fluid and use a linear $f(Q)$ model
$f(Q)=\zeta Q$, where $\zeta$ is a non-zero constant. We derive
exact relativistic solutions of the corresponding field equations.
Furthermore, we study its geometric and physical properties through
astrophysical observations from the pulsar SAX J1748.9-2021, which
is found in X-ray binary systems within globular clusters. We
examine features like anisotropic pressure, the mass-radius
relationship, redshift, the Zeldovich condition, energy and
causality conditions, the adiabatic index, the
Tolman-Oppenheimer-Volkoff equation, the equation of state parameter
and compactness. Our findings align with the observational data
which indicate that the pulsar SAX J1748.9-2021 is viable and stable
under this modified theory of gravity.
\end{abstract}
\textbf{Keywords}: Stellar configurations; Pulsar; $f(Q)$ gravity.\\
\textbf{PACS}: 97.10.-q; 97.60.Gb; 04.50.Kd.

\section{Introduction}

In the early 20th century, Albert Einstein introduced the general
theory of relativity (GR), which revolutionized our understanding of
the cosmos. This theory has been supported by numerous accurate
observations, helping us to uncover many hidden aspects of the
cosmos in modern cosmology. Subsequent observations of supernovae
have confirmed that our universe is presently experiencing a rapid
expansion phase \cite{1}. There is a substantial evidence indicating
that our universe is largely influenced by the mysterious components
known as dark matter and dark energy (DE). Identifying the unknown
form of energy is a challenging task for modern researchers. In GR,
the cosmological constant $\Lambda$ is the simplest way to explain
vacuum energy \cite{2}. However, this approach has its limitations,
as it cannot solve issues like fine-tuning \cite{3} and coincidence
problems \cite{4}. In this context, it is believed that GR might not
be the best model for describing gravity on large scales. Despite
limited progress in understanding cosmic acceleration, research into
modified theory of gravity (MTG) remains essential. These studies
provide strong and logical alternatives to GR and might solve some
current issues. Over the past two decades, numerous studies have
been conducted on MTGs to explore and better understand the
structure of the cosmos \cite{6}.

General relativity describes gravitational interactions using the
Levi-Civita connection in Riemannian spacetime. This approach relies
on the assumption of a geometry which is free from torsion and
non-metricity. Additionally, it is important to remember that the
general affine connection can be expressed more broadly \cite{7}.
Teleparallel gravity is an alternative theory to GR that explains
gravitational interactions and is characterized by torsion $T$
\cite{8}. Its teleparallel equivalent of GR (TEGR) uses the
Weitzenb$\ddot{o}$ck connection, which means there is zero curvature
and non-metricity involved \cite{9}. In a cosmological model within
Weyl-Cartan spacetime, the Weitzenb$\ddot{o}$ck connection involves
the concept that the sum of the curvature and the scalar torsion
vanishes \cite{10}. The Riemann-Cartan spacetime is similar to the
TEGR when the non-metricity is zero.

Symmetric teleparallel gravity is another alternative theory that
assumes zero curvature and torsion while including non-metricity $Q$
\cite{11}. The extension of these concepts leads us to $f(Q)$
gravity \cite{12}. This theory has gained attention in research
\cite{13}, exploring various geometric and physical implications,
providing a new perspective on gravity and cosmic phenomena. Lazkoz
et al. \cite{14} examined the constraints of $f(Q)$ gravity using
polynomial expressions related to redshift and analyzed energy
conditions for two distinct models within this framework. Shekh
\cite{15} performed a dynamic analysis of the holographic DE model
in the same gravity. Frusciante \cite{16} proposed a model of this
gravity that shares foundational similarities with the standard
$\Lambda$ CDM model. In recent papers \cite{17}, we have constructed
the generalized ghost DE and generalized ghost pilgrim DE models in
$f(Q)$ using the correspondence principle within a non-interacting
framework. We have also studied the pilgrim and generalized ghost
pilgrim DE models for the non-interacting case \cite{18}. These
models reproduce different cosmic epochs, exploring a phantom phase
of the universe and are consistent with the latest observational
data.

Research on stellar structures within MTGs has gained significant
attention over the past few decades. Pulsar SAX J1748.9-2021 (PS) is
a neutron star located in our galaxy, known for its rapid rotation
and strong magnetic field. Discovered in 1998 by the BeppoSAX/WFC
satellite \cite{27}, it has emitted beams of radiation that create
detectable pulses as it spins. Since its discovery, PS has
experienced four outbursts, occurring approximately every five years
\cite{28}, with the last one in 2015 \cite{31}. These events help
scientists to study the pulsar's behavior and the extreme conditions
present in neutron stars. An interesting aspect of PS is its role in
an X-ray binary system, where it is paired with a companion star.
Observations of these X-rays provide valuable information about the
pulsar mass and structure. Research on PS enhances our understanding
of neutron stars and the fundamental physics that govern these
intriguing celestial objects \cite{32}.

Pulsars might not have a direct impact on cosmology, but they are
very important for studying gravitational waves, researching dark
matter, measuring cosmic distances accurately and other areas of
precision astronomy. They are especially important for testing MTGs.
Kramer et al. \cite{7a} studied the double pulsar system PSR
J0737-3039A/B and highlighted its exceptional potential for
conducting precise tests of GR and alternative theories. Its unique
characteristics and proximity suggest that it could provide more
accurate tests than current solar system experiments and may
challenge existing assumptions about pulsar formation. Nashed
\cite{7d} studied matter and geometry coupling in millisecond
pulsars in $f(\mathcal{R},\mathbb{T})$ theory
($\mathcal{R},\mathbb{T}$ denote as Ricci scalar and trace of energy
momentum tensor (EMT), respectively). They found that
matter-geometry interactions lead to smaller star sizes as compared
to GR and observations of 22 pulsars supported this theory, which
fits well with linear trends and allows for neutron star mass up to
3.35 times the Sun mass. Recent studies have explored the effects of
$f(\mathcal{R})$ and $f(\mathcal{R},\mathbb{T})$ gravity theories on
PS in X-ray binary systems \cite{36}.

Maurya and Gupta \cite{7b} extended their model to derive solutions
for an anisotropic charged fluid distribution, demonstrating that
anisotropy and electric intensity increase from the core to the
surface. Sharif and Waseem \cite{1bb} examined the effects of charge
on anisotropic relativistic compact star candidates within the
$f(\mathcal{R},\mathbb{T})$ gravity. Using data from Her X-1,
4U1820-30 and SAX J 1808.4-3658, they explored physical properties
and assessed stability, concluding that charge enhances stability.
Sharif and Gul \cite{{1dd}} explored the properties of charged
compact stars with anisotropic matter using the embedding class-1
technique in $f(\mathcal{R},\mathbb{T}^2)$ theory. Bhattacharjeea
and Chattopadhyaya \cite{4aa} explored spherically symmetric,
anisotropic charged compact stars within the $f(Q)$ gravity
framework, using the Krori-Barua metric and a linear $f(Q)$ model.
They found that the charge intensity affects the star mass and
radius and the model satisfies all stability as well as physical
criteria, aligning with observational data.

This paper explores the impact of charge on the viable features and
stability of anisotropic PS within the context of $f(Q)$ gravity.
The structure of the paper is as follows. Section \textbf{2}
explains the fundamentals of $f(Q)$ gravity and its field equations
with charge. It also covers how the Einstein-Maxwell equations are
formulated and solved within this theory. In section \textbf{3}, we
present the field equations for a specific $f(Q)$ model and apply
the KB ansatz. Matching conditions are used to determine the unknown
constants in the KB ansatz. Sections \textbf{4} and \textbf{5}
utilize observational data from PS to derive density, radial and
tangential pressures, anisotropy, the mass-radius relation,
redshift, the Zeldovich condition, energy conditions, the causality
condition, the adiabatic index, the Tolman-Oppenheimer-Volkoff (TOV)
equation, the equation of state (EoS) parameter and compactness. We
also evaluate the stability of the model based on these physical
constraints. Finally, our conclusions are presented in section
\textbf{6}.

\section{$f(Q)$ Gravity and the Einstein-Maxwell Equations}

The action for $f(Q)$ gravity, which incorporates the matter
Lagrangian $L_{m}$ and the Lagrangian of the electromagnetic field
$L_{e}$, is expressed as follows \cite{12}
\begin{equation}\label{1}
S=\int\frac{1}{2}f(Q) \sqrt{-g}d^{4}x+\int(L_{m}+L_{e})
\sqrt{-g}d^{4}x.
\end{equation}
The Lagrangian for the electromagnetic field is given by
\begin{equation}\label{2}
L_{e}=-\frac{1}{16\pi} F_{\alpha\beta} F^{\alpha\beta},
\end{equation}
where $ F_{\alpha\beta} = \varphi_{\alpha,\beta} -
\varphi_{\beta,\alpha} $ represents the Maxwell field tensor and
$\varphi_{\alpha}$ denotes the four-potential. The determinant of
the metric tensor is defined as $g$, $L_{m}$ is the matter
Lagrangian density. The non-metricity scalar is described as
\begin{equation}\label{3}
Q=-g^{\mu\nu}(\mathbb{L}^{\varrho}_{~\phi\mu}\mathbb{L}^{\phi}_{~\nu\varrho}
-\mathbb{L}^{\varrho}_{~\phi\varrho}\mathbb{L}^{\phi}_{~\mu\nu}),
\end{equation}
where
\begin{equation}\label{4}
\mathbb{L}^{\varrho}_{\;\phi\mu}=-\frac{1}{2}g^{\varrho\lambda}
(\nabla_{\mu}g_{\phi\lambda}+\nabla_{\phi}g_{\lambda\mu}
-\nabla_{\lambda}g_{\phi\mu}).
\end{equation}
The superpotential is characterized as
\begin{equation}\label{5}
\mathbb{P}^{\varrho}_{\;\mu\nu}=-\frac{1}{2}\mathbb{L}^{\varrho}_{\;\mu\nu}
+\frac{1}{4}(Q^{\varrho}-\tilde{Q}^{\varrho})g_{\;\mu\nu}-
\frac{1}{4} \delta ^{\varrho}\;_{({\mu}}Q_{\nu)},
\end{equation}
where $Q_{\varrho}=Q^{~\mu}_{\varrho~\mu}$ and
$\tilde{Q}_{\varrho}=Q^{\mu}_{~\varrho\mu}$. Thus, the expression
for $Q$ becomes \cite{32-a}
\begin{equation}\label{6}
Q=-Q_{\varrho\zeta\tau}\mathbb{P}^{\varrho\zeta\tau}=
-\frac{1}{4}(-Q^{\varrho\tau\rho}Q_{\varrho\tau\rho}
+2Q^{\varrho\tau\rho}Q_{\rho\varrho\tau}-2Q^{\rho}
\tilde{Q}_{\rho}+Q^{\rho}Q_{\rho}).
\end{equation}
The field equations associated with $f(Q)$ gravity can be written as
\begin{equation}\label{7}
\frac{-2}{\sqrt{-g}}\nabla_{\varrho}(f_{Q}\sqrt{-g}
P^{\varrho}_{~\mu\nu})-\frac{1}{2}f g_{\mu\nu}-f_{Q}
(P_{\mu\varrho\phi}Q_{\nu}^{~\varrho\phi}-2Q^{\varrho\phi}_{~~~\mu}
P_{\varrho\phi\nu})=\mathbb{T}_{\mu\nu}+E_{\mu\nu},
\end{equation}
here $f_{Q}$ represents the derivative with respect to $Q$.

The line element for a spherically symmetric spacetime is
\begin{equation}\label{8}
ds^{2}=e^{\xi(r)}dr^{2}-e^{\psi(r)}dt^{2}
+r^{2}(d\theta^{2}+\sin^{2}\theta d\phi^{2}).
\end{equation}
An anisotropic perfect fluid is a type of fluid in which the
pressure differs in different directions, unlike an isotropic
perfect fluid where the pressure is same in all directions. The EMT
for an anisotropic perfect fluid can be written as
\begin{equation}\label{9}
\mathbb{T}_{\mu\nu}=(\rho+p_{t})u_{\mu}u_{\nu}+p_{t}g_{\mu\nu}-\chi
k_{\mu}k_{\nu},
\end{equation}
In this context, $\rho$ represents the energy density of the fluid,
while $p_r$ is the radial pressure and $p_t$ is the tangential
pressure. The four-velocity components of the fluid, denoted as
$u_{\mu}$ and $u_{\nu}$, adhere to the normalization condition
$u^{\mu}u_{\mu} = -1$. Additionally, $k_{\mu}$ is a unit four-vector
in the radial direction, satisfying $k^{\mu}k_{\mu} = 1$. The
anisotropy factor $\chi$ is defined as the difference between the
tangential pressure and the radial pressure, expressed as
$\chi=p_t-p_r$. Using the provided metric, $u^\mu$ can be written as
$e^{-\psi} \delta^\mu_t$ and $k^\mu$ as $e^{-\xi} \delta^\mu_r$. As
a result, the trace of the EMT \eqref{9} can be written as
\begin{equation}\label{10}
\mathbb{T}=-\rho+3p_{r}+2\chi.
\end{equation}

The stress-energy tensor for the electromagnetic field is expressed
by
\begin{equation}\label{11}
E_{\mu\nu} = \frac{1}{4\pi} \bigg( F^{\beta}_\mu F_{\nu\beta} -
\frac{1}{4} g_{\mu\nu} F_{\alpha\beta} F^{\alpha\beta} \bigg).
\end{equation}
The corresponding Maxwell equations are
\begin{equation}\label{12}
(\sqrt{-g} F_{\mu\nu})_{;\nu} = 4\pi J_\mu \sqrt{-g}, \quad
F_{[\mu\nu;\delta]} = 0.
\end{equation}
In this context, it specifically refers to the covariant derivative
with respect to the Levi-Civita connection, which is the standard
connection used in GR for ensuring that the covariant derivative is
compatible with the metric and has no torsion. Where $J_\mu = \sigma
u_\mu$ represents the electric four-current, $\sigma$ is the charge
density. The electric field strength $E(r)$ is expressed as
\begin{equation}\label{13}
E(r) = \frac{e^{\frac{\psi + \xi}{2}}}{r^2} q(r),
\end{equation}
where $q(r)$ represents the total charge enclosed within a sphere of
radius $r$, calculated as
\begin{equation}\label{14}
q(r) = 4\pi \int_0^r \sigma r^2 e^\xi dr,
\end{equation}
leading to
\begin{equation}\label{15}
\sigma = \frac{e^{-\xi}}{4\pi r^2} \frac{dq(r)}{dr}.
\end{equation}
In this study, we use a specific charge distribution given by
\cite{2bb}
\begin{equation}\label{16}
q = q_0 r^3,
\end{equation}
where $q_0$ represents the charge intensity. If $q_0 = 0$, it
corresponds to a neutral, uncharged scenario. The field equations
\eqref{7} have the following non-zero components
\begin{eqnarray}\nonumber
\rho&=&\frac{1}{2 r^2 e^{\xi}}\bigg[f_{Q} \bigg(\big(e^{\xi}-1\big)
\big(r \psi^{\prime}+2\big)+r \big(e^{\xi}+1\big)
\xi^{\prime}\bigg)\\
\label{20}&+&2rf_{QQ}\big(e^{\xi}-1\big)Q^{\prime}+f r^2
e^{\xi}\bigg]-\frac{(q_0 r^3)^2}{r^4}, \\\nonumber
p_{r}&=&\frac{1}{2 r^2 e^{\xi}}\bigg[f_{Q}\bigg(\big(e^{\xi}-1\big)
\big(r \psi^{\prime}+r \xi^{\prime}+2\big)-2 r
\psi^{\prime}\bigg)\\
\label{21}&+&2f_{QQ}r \big(e^{\xi}-1\big)Q^{\prime}+fr^2
e^{\xi}\bigg]+\frac{(q_0 r^3)^2}{r^4}, \\\nonumber
p_{t}&=&\frac{-1}{4 r e^{\xi}}\bigg[f_{Q} \big(\xi^{\prime} \big(r
\psi^{\prime}+2 e^{\xi}\big)-r
\psi^{\prime2}-2\psi^{\prime\prime}r+2
\big(e^{\xi}-2\big) \psi^{\prime}\big)\\
\label{22}&-&2 f_{QQ} r Q^{\prime} \psi^{\prime}+2fr
e^{\xi}\bigg]-\frac{(q_0 r^3)^2}{r^4},
\end{eqnarray}
where prime is the derivative with respect to $r$. The non-metricity
scalar can be given as
\begin{equation}\label{23}
Q=\frac{e^{-\xi}
\big(1-e^{\xi}\big)\big(\psi^{\prime}+\xi^{\prime}\big)}{r}.
\end{equation}

\section{The $f(Q)$ Gravity Model}

The most widely recognized initial model of $f(Q)$ theory is the
linear model \cite{36c}
\begin{equation}\label{24}
f(Q)=\zeta Q,
\end{equation}
where $\zeta$ is an arbitrary constant and it is not fixed like
gravitational constant $G$ but is instead varied using a hit and
trail approach to optimize the model for different cosmic behaviors.
Thus, it allows flexibility in matching the model to observational
data by adjusting the parameter to find the best-fit behavior of key
cosmic parameters like expansion rates or DE contributions. As this
constant is determined theoretically, it differs from the $G$ in
terms of its role in the theory. While $G$ is universally defined
across all gravitational phenomena, the modified gravitational
coupling in this model is treated more as a tunable parameter to
accommodate specific cosmic observations. Therefore, the comparison
between the two is limited, as the arbitrary constant does not
universally define gravitational strength but is used to explore how
varying this constant affects the behavior of the STEGR model,
especially in regimes where deviations from GR are expected. In
particular, the linear model clarifies that when $\zeta=1$, the
model reduces to STEGR, fully recovering the gravitational dynamics
of GR. Here $f_{Q}=\zeta$ and $f_{QQ}=0$. We substitute these values
into Eqs.\eqref{20}-\eqref{22} and obtain
\begin{eqnarray}\label{25}
\rho&=&\frac{-q^2+\zeta  r^2 e^{-\xi (r)} \bigg(r
\xi^{\prime}-1\bigg)+\zeta  r^2}{r^4}, \\\nonumber
p_{r}&=&\frac{1}{2 r^2}\bigg[\big(2 r^3 e^{2 \xi} \big(-\zeta
\big(\psi^{\prime}+\xi^{\prime}\big)\big)+\zeta e^{\xi} \big(r
\big(2 r^2+1\big)
\big(\psi^{\prime}+\xi^{\prime}\big)+2\big)\\
\label{26}&-&\zeta \big(3 r \psi^{\prime}+r
\xi^{\prime}+2\big)\big)e^{-\xi}\bigg]+\frac{q^2}{r^4},\\
\label{27} p_{t}&=&\frac{1}{4}\bigg[\frac{1}{r}\bigg\{\zeta e^{-\xi}
\bigg(r \psi ^{\prime2}-2 \xi^{\prime}+2 r
\psi^{\prime\prime}+\psi^{\prime} \big(2-r
\xi^{\prime}\big)\bigg)\bigg\}\bigg]-\frac{q^2}{r^4},
\end{eqnarray}
To understand how stars evolve, we make some reasonable assumptions
about two key functions $\xi(r)$ and $\psi(r)$.

\subsection{The Krori-Barua Spacetime}

The KB scheme is a set of metric potentials used to model compact
objects in GR as well as in various MTGs. This scheme provides a
framework for describing the gravitational field within a
spherically symmetric star \cite{37}. The metric potentials in the
KB spacetime are expressed as
\begin{equation}\label{28} \psi(r)=\eta_{0}
r^2+\eta_{1},~~~ \xi(r)=\eta_{2}r^2,
\end{equation}
where $\eta_{0}, \eta_{1},\eta_{2}$ are arbitrary constants
determined by junction conditions. This solution ensures that the
interior of the star is smooth and free of singularities.

\subsection{Matching Conditions}

In a stellar system, we can assume that solutions from
non-Riemannian geometry in a vacuum are the same as those from
$f(Q)$ theory. This means that the model described by \eqref{24}
applies here as well. As a result, the external solution aligns with
the vacuum charged exterior Reissner-Nordstr$\ddot{o}$m (RN)
solution. Therefore, we use the following model to describe the
exterior spacetime as
\begin{equation}\label{29}
ds^{2}=-\bigg(1+\frac{\mathbb{Q}^{2}}{r^{2}}-\frac{2\mathbb{M}}{r}\bigg)dt^{2}
+\bigg(1+\frac{\mathbb{Q}^{2}}{r^{2}}-\frac{2\mathbb{M}}{r}\bigg)^{-1}dr^{2}
+r^{2}(d\theta^{2}+\sin^{2}\theta d\phi^{2}).
\end{equation}
Here $\mathbb{M}$ denotes the gravitational mass of the star.
Ensuring that the metric coefficients from \eqref{8}, \eqref{28} and
\eqref{29} are continuous at the boundary surface ($r=\mathbb{R}$)
leads to the geometric component given by
\begin{eqnarray}\label{1a}
g_{tt}&=&e^{\eta_{0}\mathbb{R}^2
+\eta_{1}}=\bigg(1+\frac{\mathbb{Q}^{2}}{\mathbb{R}^{2}}-\frac{2\mathbb{M}}{\mathbb{R}}\bigg),\\
\label{1b} g_{rr}&=&e^{\eta_{2}\mathbb{R}^2}=
\bigg(1+\frac{\mathbb{Q}^{2}}{\mathbb{R}^{2}}-\frac{2\mathbb{M}}{\mathbb{R}}\bigg)^{-1},\\
\label{1c}g_{tt,r}&=&2\eta_{0}\mathbb{R} e^{\eta_{0}\mathbb{R}^2
+\eta_{1}}=\frac{2\mathbb{M}}{\mathbb{R}^2}-\frac{2\mathbb{Q}^2}{\mathbb{R}^3}.
\end{eqnarray}
Solving \eqref{1a}-\eqref{1c} and applying the condition
$p_{r}(r=\mathbb{R})=0$, the constants $\eta_{0}$, $\eta_{1}$ and
$\eta_{2}$ can be found as follows
\begin{eqnarray}\label{2a}
\eta_{0}&=&
\bigg(-\frac{2\mathbb{Q}^2}{\mathbb{R}^4}+\frac{2\mathbb{M}}{\mathbb{R}^3}\bigg)\bigg(-\frac{2\mathbb{M}}{\mathbb{R}}+1
+\frac{\mathbb{Q}^{2}}{\mathbb{R}^{2}}\bigg)^{-1},\\
\label{2b}\eta_{1}&=&\ln\bigg(1-\frac{2\mathbb{M}}{\mathbb{R}}+\frac{\mathbb{Q}^{2}}{\mathbb{R}^{2}}\bigg)
-\bigg(\frac{2\mathbb{M}}{\mathbb{R}^3}-\frac{2\mathbb{Q}^2}{\mathbb{R}^4}\bigg)\bigg(1-\frac{2\mathbb{M}}{R}
+\frac{\mathbb{Q}^{2}}{\mathbb{R}^{2}}\bigg)^{-1},\\
\label{2c} \eta_{2}&=&\frac{\ln \bigg(\frac{\mathbb{R}^2}{-2
\mathbb{M}
\mathbb{R}+\mathbb{Q}^2+\mathbb{R}^2}\bigg)}{\mathbb{R}^2}.
\end{eqnarray}

\section{Stability Analysis from SAX J1748.9-2021 Observations}

In this section, we utilize observational data, specifically the
mass and radius of the PS within the $f(Q)$ gravity. We examine the
stability of the solution to be obtained by applying various
constraints. Accurate observational data is essential for
determining the model parameters. We rely on detailed spectroscopic
data from EXO 1745-248 during thermonuclear bursts, which has given
precise measurements for the pulsar mass ($\mathbb{M}=1.81\pm
0.3\mathbb{M}_{\odot}$) and radius ($\mathbb{R}=11.7 \pm 1.7 km$)
\cite{37a}. Using these measurements, we explore physical
characteristics and stability of the pulsar.

\subsection{The Material Component}

Substituting Eq.\eqref{28} into \eqref{25}-\eqref{27}, we can derive
the expressions for $\rho$, $p_{r}$ and $p_{t}$ as follows
\begin{eqnarray}\label{32}
\rho&=&\frac{\zeta  e^{-(\eta_{2} r^2)} \bigg(2 \eta_{2}
r^2-1\bigg)+\zeta -q_{0}^2 r^4}{r^2},
\\\nonumber p_{r}&=&\frac{1}{r^2}\bigg[e^{-\eta_{2} r^2}
\bigg(e^{\eta_{2} r^2} \bigg(\zeta  \big(\big(2 r^2+1\big) r^2
(\eta_{0}+\eta_{2})\big)+q_{0}^2 r^4\bigg)\\\nonumber &-&\zeta
\big(r^2 (3 \eta_{0}+\eta_{2})\big)-2 \zeta  r^4 (\eta_{0}+\eta_{2})
e^{2 \eta_{2} r^2}\bigg)\bigg],
\\\label{34} p_{t}&=&\zeta e^{-\eta_{2} r^2}
\bigg(\eta_{0} r^2 (\eta_{0}-\eta_{2})(2
\eta_{0}-\eta_{2})\bigg)-q_{0}^2 r^2.
\end{eqnarray}
We create plots of $\rho$, $p_r$ and  $p_t$ against the radial
distance. For our analysis, we have used the values
$\eta_{0}=0.00363$, $\eta_{1}=-0.40562$ and $\eta_{2}=0.00405$. We
choose $Q=3$ because it gibes better results in the graphs.
Additionally, we vary the value of $q_{0}$ as $0.004$, $0.006$ and
$0.008$ for our graphical analysis. Higher $q_{0}$ increases
repulsion from the electric field, which affects the gravitational
potential and makes the charge's effect more visible. In $f(Q)$
theory, smaller charges result in fewer changes in the gravitational
field, whereas larger charges lead to stronger electric repulsion.
The plots of $\rho$, $p_r$ and $p_t$ are shown in Figure \textbf{1}.
These values are highest at the center of the star and gradually
decrease as we move outward. This shows that the matter is very
dense at the center and becomes less dense towards the edge,
indicating a highly concentrated profile of the PS with increasing
$r$.
\begin{figure}\center
\epsfig{file=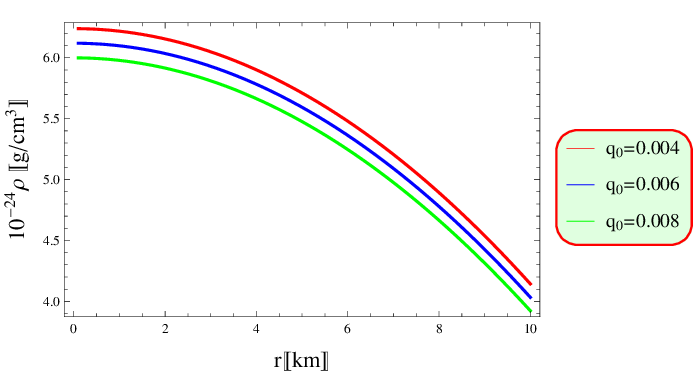,width=.47\linewidth}
\epsfig{file=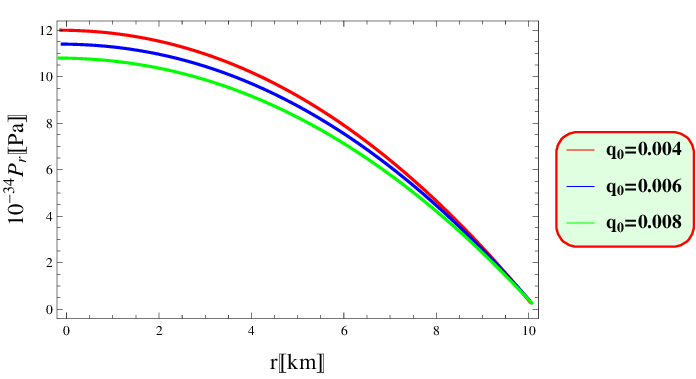,width=.47\linewidth}
\epsfig{file=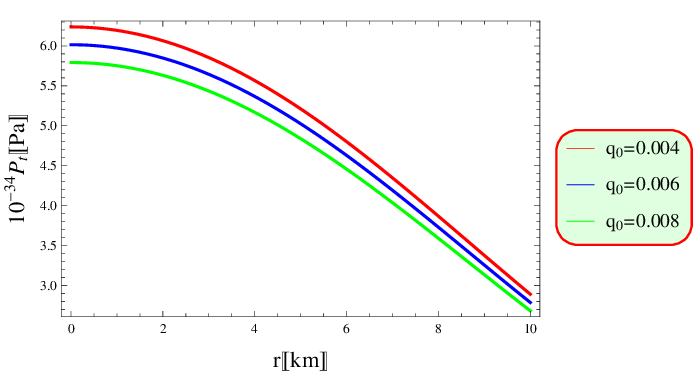,width=.47\linewidth}\caption{Graphs of $\rho$,
$p_{r}$, $p_{t}$ as functions of $r$.}
\end{figure}

Anisotropic pressure refers to the condition where the pressure
inside a star or similar object varies in different directions. This
can happen because of quantities like strong magnetic fields, the
star rotation or different types of matter inside the star. This
variation in pressure can have a big impact on the stability and
structure of compact objects like PS \cite{38}. It is expressed as
\begin{eqnarray}\nonumber
\chi &=&\frac{1}{r^2}\bigg[e^{-\eta_{2} r^2} \bigg(\zeta
\big(\eta_{0} \eta_{2} r^4-\eta_{0}^2 r^4-5 \eta_{0} r^2\big)+
e^{\eta_{2} r^2}\bigg(2 q_{0}^2 r^4+\zeta \big(2 \eta_{0}
r^4\\\nonumber &+&\eta_{0} r^2+2 \eta_{2} r^4+\eta_{2}
r^2\big)\bigg)-2 \zeta  r^4 (\eta_{0}+\eta_{2}) e^{2 \eta_{2}
r^2}\bigg)\bigg].
\end{eqnarray}
Figure \textbf{2} demonstrates that the anisotropy meets the
stability criteria. It begins from zero at the center of the star
and gradually increases in a consistent manner towards the star's
surface.

Providing numerical values for the physical properties of the pulsar
as predicted by the current model, is crucial. For example, when
$q_{0}$ is 0.004, $\rho_{core}\approx 6.24 \times 10^{16}$
g/cm$^{3}$, $p_{r(core)}\approx 11.95 \times 10^{34}$ dyn/cm$^{2}$
and $p_{t(core)}\approx 6.24 \times 10^{34}$ dyn/cm$^{2}$. At the
star's surface, $\rho_{I}\approx 4.14 \times 10^{16}$ g/cm$^{3}$
with $p_{r(r=\mathbb{R})}= 0$ dyn/cm$^{2}$ and
$p_{t(r=\mathbb{R})}\approx 2.89 \times 10^{34}$ dyn/cm$^{2}$. For
$q_{0}=0.006$, $\rho_{core}\approx 6.12 \times 10^{16}$ g/cm$^{3}$,
$p_{r(core)}\approx 11.31 \times 10^{34}$ dyn/cm$^{2}$ and
$p_{t(core)}\approx 5.99 \times 10^{34}$ dyn/cm$^{2}$. At the star's
surface, $\rho_{I}\approx 4.06 \times 10^{16}$ g/cm$^{3}$ with
$p_{r(r=\mathbb{R})}=0$ dyn/cm$^{2}$ and $p_{t(r=\mathbb{R})}\approx
2.78 \times 10^{34}$ dyn/cm$^{2}$. When $q_{0}=0.008$,
$\rho_{core}\approx 5.98 \times 10^{16}$ g/cm$^{3}$,
$p_{r(core)}\approx 10.79 \times 10^{34}$ dyn/cm$^{2}$ and
$p_{t(core)}\approx 5.79 \times 10^{34}$ dyn/cm$^{2}$. At the star's
surface, $\rho_{I}\approx 3.94 \times 10^{16}$ g/cm$^{3}$ with
$p_{r(r=\mathbb{R})}=0$ dyn/cm$^{2}$ and $p_{t(r=\mathbb{R})}\approx
2.68 \times 10^{34}$ dyn/cm$^{2}$.
\begin{figure}\center
\epsfig{file=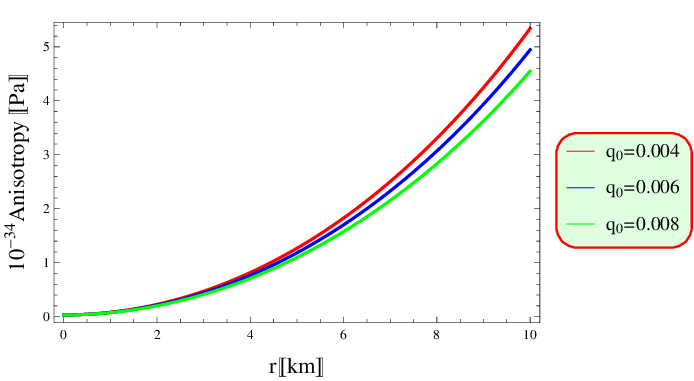,width=.5\linewidth} \caption{Graphical
representation of $\chi$ relative to $r$.}
\end{figure}

\subsection{Limit on the Mass-Radius Relation}

The mass of PS is given as \cite{37a}
\begin{equation}\nonumber
M(r)=4\pi\int_0^r\gamma^2\rho(\gamma)d\gamma.
\end{equation}
Using Eq.\eqref{32} and performing the integration, we obtain
\begin{equation}\label{35}
M(r)= 4 \pi  \bigg(\zeta  r \bigg(1-e^{-\eta_{2}
r^2}\bigg)-\frac{q_{0}^2 r^5}{5}\bigg).
\end{equation}
This numerical analysis evaluates the reliability of the current
model, as depicted in Figure \textbf{3}. The mass function rises
steadily and evenly as the star's radius increases.
\begin{figure}\center
\epsfig{file=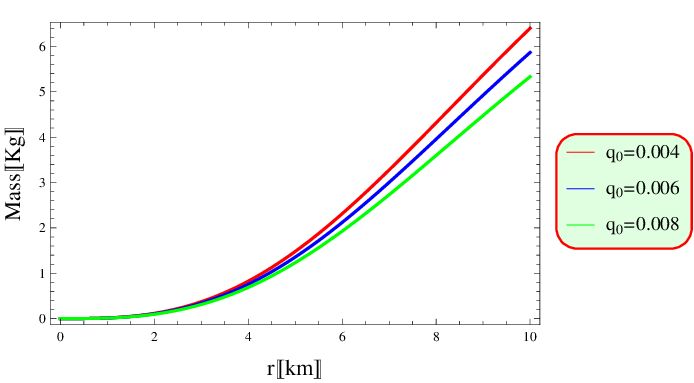,width=.5\linewidth} \caption{Graph of
mass-radius as functions of $r$.}
\end{figure}

\subsection{The Gravitational Redshift}

This is the effect where light or electromagnetic radiation shifts
to longer wavelengths as it escapes from a strong gravitational
field. This occurs because the energy of the photons decreases in
the gravitational potential of a massive object. The stronger the
gravitational field, such as near a black hole or a dense star, the
more significant the redshift. This phenomenon, predicted by GR,
helps us to understand the influence of gravity on light and can be
observed in the spectra of stars and other celestial objects. Ivanov
\cite{41} calculated the value to be 5.211 for anisotropic
configurations that adhere to the dominant energy condition. The
gravitational redshift function is given by
\begin{equation}\nonumber
Z(r)=\frac{1}{\sqrt{1-\varpi}}-1,
\end{equation}
where $\varpi= \frac{2 M}{r}$. Applying Eq.\eqref{35}, we obtain
\begin{equation}\label{35a}
Z(r)=\frac{1}{\sqrt{8 \pi  \zeta \bigg(e^{-\eta_{2}
r^2}-1\bigg)+\frac{8}{5} \pi  q_{0}^2 r^4+1}}-1.
\end{equation}
We create a graph of the redshift function for the pulsar using
different values of $q_{0}$. The gravitational redshift stays
positive and finite inside the star and gradually increases, as
shown in Figure \textbf{4} \cite{42}. The redshift function stays
within the specified limit ($Z(r)<5.211$).
\begin{figure}\center
\epsfig{file=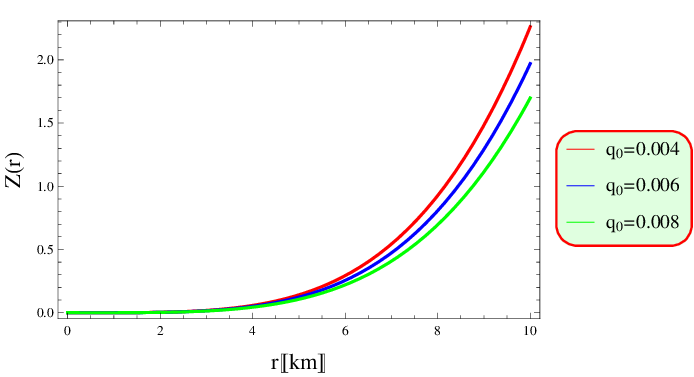,width=.5\linewidth} \caption{Graph of redshift
as functions of $r$.}
\end{figure}

\subsection{The Zeldovich Condition}

The Zeldovich condition \cite{43} is a key criterion in astrophysics
for evaluating the stability of stars, particularly within the
frameworks of GR and stellar astrophysics. According to the
Zeldovich condition, the ratio of the central pressure $p(0)$ to the
central energy density $\rho(0)$ must satisfy specific requirements
to ensure the stability of the star. This condition helps in
understanding the balance between gravitational forces and the
internal pressure that prevents collapse. It is defined as
\begin{equation}\label{36}
\frac{p(0)}{c^2 \rho(0)} \leq 1.
\end{equation}
We should verify that this ratio does not exceed 1. The values of
$\rho$, $p_{r}$ and $p_{t}$ when $r\rightarrow 0$ can be determined
as follows
\begin{equation}\nonumber
\rho(0)=2\eta_{2} \zeta,\quad p_{r}(0)=\zeta (-2 \eta_{0} +
\eta_{2}),\quad p_{t}(0)=(2\eta_{0} - \eta_{2}) \zeta.
\end{equation}
Using the values previously calculated for the PS in section
\textbf{4.1}, we can assess Zeldovich's inequality \eqref{36}.
According to the expressions above, the ratio
$\frac{p_{r}(0)}{\rho(0)}$ is 0.7, which is less than 1. Similarly,
the ratio $\frac{p_{t}(0)}{\rho(0)}$ is -1.1, also less than 1.
These results show that the Zeldovich condition is satisfied in both
cases.

\subsection{Energy Conditions}

Energy conditions are important criteria in GR and cosmology, used
to constrain the behavior of matter and energy in spacetime. They
help to ensure that the EMT behaves in a physically reasonable way.
There are several types of energy conditions
\begin{enumerate}
\item Null energy condition is represented as

$0 \leq \rho + p_r, \quad 0 \leq \rho + p_t$.
\item Dominant energy condition is expressed as

$0 \leq \rho - p_r, \quad 0 \leq \rho - p_t$.
\item Weak energy condition is presented as

$ 0 \leq \rho + p_r, \quad 0 \leq \rho + p_t, \quad 0 \leq \rho$.
\item Strong energy condition is stated as

$ 0 \leq \rho + p_r, \quad 0 \leq \rho + p_t, \quad 0 \leq \rho +
p_r + 2p_t$.
\end{enumerate}
These conditions are crucial in understanding how cosmic structures
form and stay stable within spacetime. They help to determine if a
PS can exist and remain stable. Figure \textbf{5} shows the energy
conditions for different values of $q_{0}$ and demonstrates that the
PS meets all the necessary energy conditions.
\begin{figure}\center
\epsfig{file=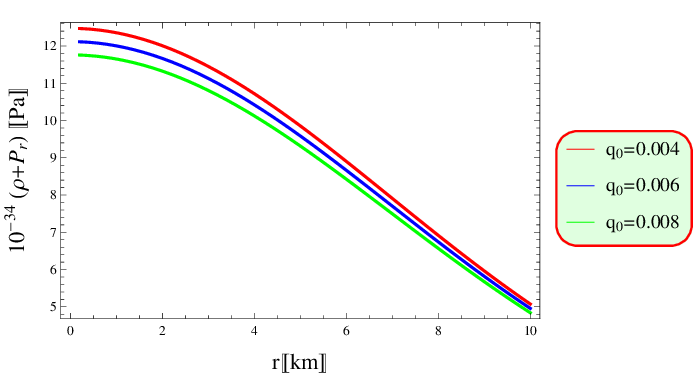,width=.5\linewidth}\epsfig{file=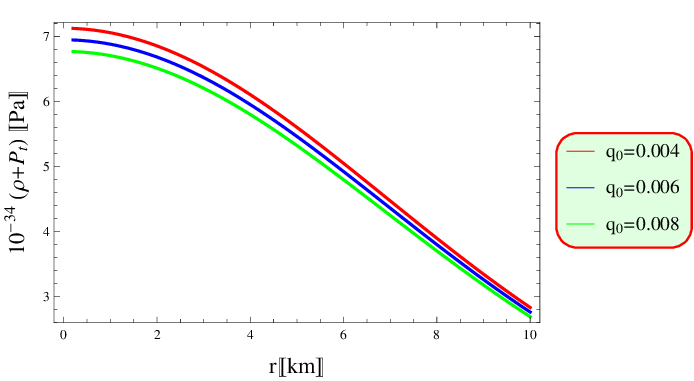,width=.5\linewidth}
\epsfig{file=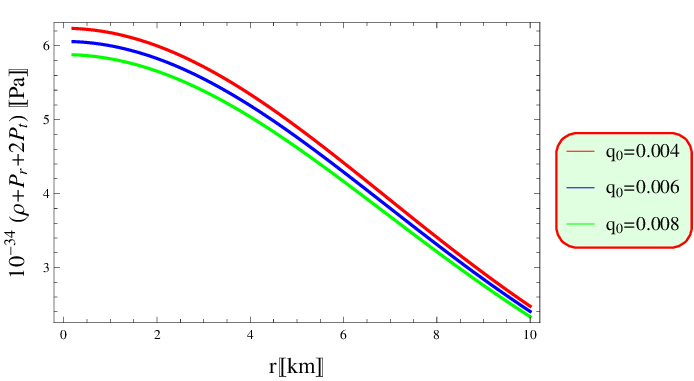,width=.5\linewidth}\epsfig{file=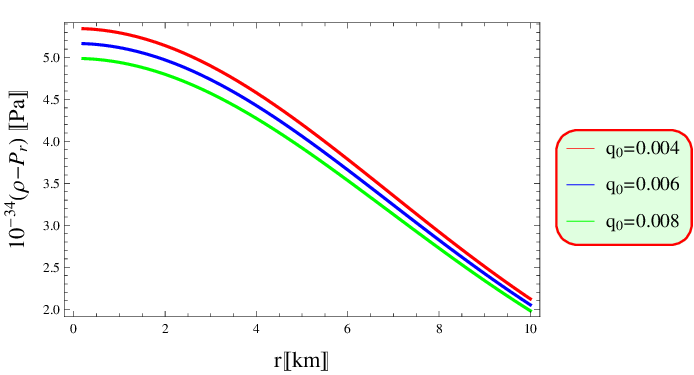,width=.5\linewidth}
\epsfig{file=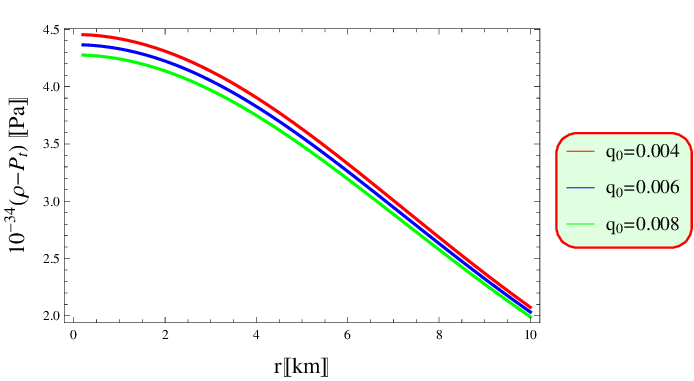,width=.5\linewidth}\caption{Graphical
representations of energy conditions relative to $r$.}
\end{figure}

\subsection{Causality Conditions}

In cosmology, these conditions ensure that nothing exceeds the speed
of light. For an anisotropic fluid, it is defined as
\begin{equation}\label{37}
v_r^2= \frac{{p}_{r}^{\prime}}{\rho^{\prime}},\quad v_t^2=
\frac{{p}_{t}^{\prime}}{\rho^{\prime}}.
\end{equation}
For a neutron star like PS, the squared speed of sound must lie
within a specific range $[0, 1]$ to maintain structural stability
\cite{44}. The radial and tangential sound speeds both satisfy the
stability criteria, which specify that $ 0 \leq \frac{v_r^2}{c^2}
\leq 1$ and $ 0 \leq \frac{v_t^2}{c^2} \leq 1 $. Additionally,
stability is assessed using the cracking condition $ 0 \leq
\frac{v_t^2 - v_r^2}{c^2} \leq 1 $ \cite{45}. If this condition is
fulfilled, the PS remains stable and can persist for a long time,
otherwise, it may collapse. Applying the equation provided, we
obtain
\begin{eqnarray}\nonumber
v_r^2&=& \bigg[\zeta  \big(2 \eta_{2}^2 r^4-\eta_{2} r^2\big)+
e^{\eta_{2} r^2} \big(\zeta +q_{0}^2 r^4\big)\bigg]\bigg[2 \zeta r^4
(\eta_{0}+\eta_{2})e^{2 \eta_{2} r^2}\big(\eta_{2}
r^2\big)-\zeta\\\label{38} &\times& \big(3 \eta_{0} \eta_{2}
r^4+\eta_{2}^2 r^4+\eta_{2} r^2 \big)- e^{\eta_{2} r^2}
\bigg(q_{0}^2 r^4 +\zeta \big(2 \eta_{0} r^4+2 \eta_{2}
r^4\big)\bigg)\bigg]^{-1},
\\\nonumber v_t^2&=& \bigg[\zeta\big(-2 \eta_{2}^2 r^4+\eta_{2} r^2
\big)- e^{\eta_{2} r^2} \big(\zeta +q_{0}^2 r^4\big)\bigg] \bigg[r^4
\bigg(\zeta  \big(\eta_{2}^2+\eta_{0} \eta_{2} \big(\eta_{2} r^2
-3\big) \\\label{39} &+& \eta_{0}^2 \big(-\eta_{2}
r^2\big)\big)-q_{0}^2  e^{\eta_{2} r^2}\bigg)\bigg]^{-1},
\end{eqnarray}
where
\begin{eqnarray}\nonumber
\rho^{\prime}&=&\frac{1}{r^2}\bigg[\big(4 \eta_{2} \zeta r
e^{-\eta_{2} r^2}\big)-\bigg(2 \eta_{2} \zeta r e^{-\eta_{2} r^2}
\bigg(2 \eta_{2} r^2-1\bigg)\bigg)-4 q_{0}^2
r^3\bigg]-\frac{1}{r^3}\\\nonumber&\times&\bigg[2 \bigg(\zeta
e^{-\eta_{2} r^2} \bigg(2 \eta_{2} r^2-1\bigg)+\zeta -q_{0}^2
r^4\bigg)\bigg],
\\\nonumber p_{r}^{\prime}&=&-\frac{1}{r}\bigg[2 \eta_{2}
e^{-\eta_{2} r^2} \bigg(e^{\eta_{2} r^2} \bigg(\zeta \big(\big(2
r^2+1\big) r^2 (\eta_{0}+\eta_{2})\big)+q_{0}^2 r^4 \bigg)-\zeta r^2
(3\eta_{0}\\\nonumber&+&\eta_{2})-2 \zeta r^4 (\eta_{0}+\eta_{2})
e^{2 \eta_{2} r^2}\bigg)\bigg]-\frac{1}{r^3}\bigg[2 e^{-\eta_{2}
r^2}\bigg(e^{\eta_{2} r^2}\big(\zeta \big(\big(2 r^2+1\big) r^2
(\eta_{0}
\\\nonumber&+&\eta_{2})\big) +q_{0}^2 r^4
\big)- \zeta\big(r^2 (3 \eta_{0}+\eta_{2})\big)-2 \zeta r^4
(\eta_{0}+\eta_{2}) e^{2 \eta_{2}
r^2}\bigg)\bigg]+\frac{1}{r^2}\bigg[e^{-\eta_{2} r^2}
\\\nonumber&\times&\bigg(\bigg(2
\eta_{2} r e^{\eta_{2} r^2} \bigg(q_{0}^2 r^4+\zeta \big(\big(2
r^2+1\big) r^2(\eta_{0}+\eta_{2})\big)\bigg)\bigg)+e^{\eta_{2} r^2}
\bigg(\zeta \big(4 r^3 (\eta_{0}
\\\nonumber&+&\eta_{2})+2\big(2 r^2+1\big) r (\eta_{0}+\eta_{2})\big)+4
q_{0}^2 r^3\bigg)-\bigg(8 \eta_{2} \zeta r^5 (\eta_{0}+\eta_{2})
e^{2 \eta_{2} r^2}\bigg)\\\nonumber&-&8 \zeta r^3
(\eta_{0}+\eta_{2}) e^{2 \eta_{2} r^2}-2 \zeta r (3
\eta_{0}+\eta_{2})\bigg)\bigg],\\\nonumber p_{t}^{\prime}&=&\big(2
\eta_{0} \zeta r (\eta_{0}-\eta_{2}) e^{-\eta_{2} r^2}\big)-2
q_{0}^2 r\big(2 \eta_{2} \zeta r \big(\eta_{0} r^2
(\eta_{0}-\eta_{2})+ (2 \eta_{0}-\eta_{2})\big)e^{-\eta_{2}
r^2}\big).
\end{eqnarray}
Figure \textbf{6} depicts how the radial and tangential squared
speed of sound vary with $r$. It shows that $v_r^2$ ranges from
0.785 to 0.793 and $v_t^2$ ranges from 0.832 to 0.849. Additionally,
the difference $(v_t^2 - v_r^2 )$ lies between 0.011 and 0.012
throughout the interior of the PS. These findings confirm that all
the necessary stability conditions are fulfilled, ensuring the
stability of the PS.
\begin{figure}\center
\epsfig{file=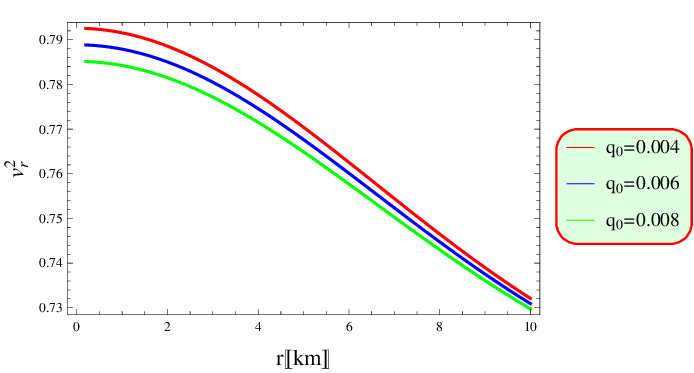,width=.5\linewidth}\epsfig{file=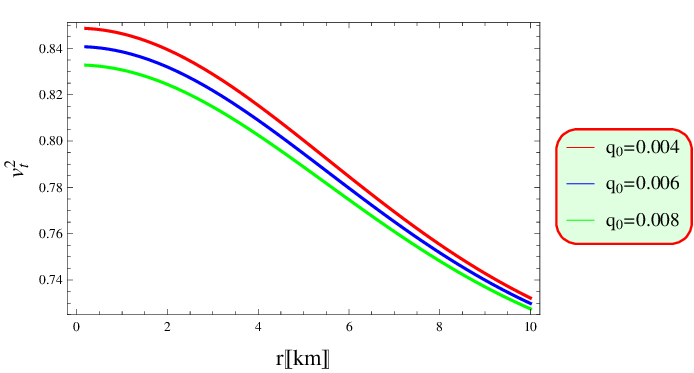,width=.5\linewidth}
\epsfig{file=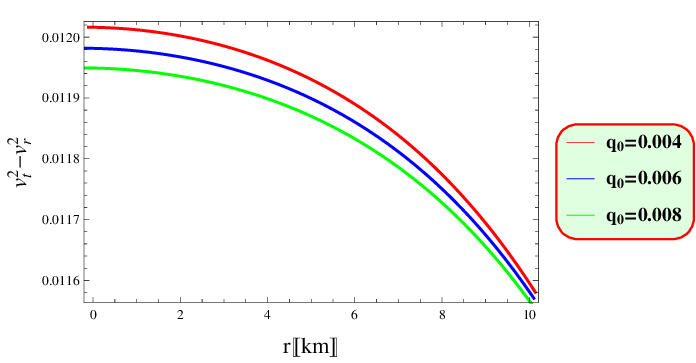,width=.5\linewidth}\caption{Graphs of causality
condition as functions of $r$.}
\end{figure}

\subsection{The Adiabatic Index and the Equilibrium of Hydrodynamic Forces}

The adiabatic index $(\Gamma)$ is a crucial parameter for assessing
the stability of stars, including PS. It provides insight into how
pressure and density affect the star's stability. It is defined by
the formula \cite{48}
\begin{equation}\nonumber
\Gamma=\frac{4}{3} \bigg(\frac{\chi}{|p_{r}^{\prime}|
r}+1\bigg)_{max}.
\end{equation}
This expression helps to determine the star's response towards the
compression and expansion which is vital for understanding its
structural integrity. For radial ($\Gamma_r$) and tangential
($\Gamma_t$) components, it is given by
\begin{equation}\nonumber
\Gamma_{r}=\frac{\nu_{r}^{2} (p_{r}+\rho c^{2} )}{p_{t}},\quad
\Gamma_{t}=\frac{\nu_{t}^{2} (p_{t}+\rho c^{2} )}{p_{t}}.
\end{equation}
For stability, $\Gamma$ must be greater than $4/3$. In an isotropic
system, where the pressure is the same in all directions $(\chi =
0)$, the parameter $\Gamma$ equals $4/3$. In cases of mild
anisotropy $(\chi < 0)$, the $\Gamma$ remains greater $4/3$, which
is consistent with the standard stability condition. In cases of
strong anisotropy $(\chi > 0)$, as explored in this study, the
$\Gamma$ can be greater than $4/3$. Neutral equilibrium is achieved
when the adiabatic indices for $\Gamma_r$ and $\Gamma_t$ are equal.
For a star to maintain stable equilibrium, both $\Gamma_r$ and
$\Gamma_t$, must be greater than the $\Gamma$. Specifically for a
PS, stability is ensured if both $\Gamma_r$ and $\Gamma_t$ are
greater than $4/3$. If these conditions are not satisfied, the star
may become unstable and could potentially collapse \cite{49}. Using
the above equation, we find
\begin{eqnarray}\nonumber
\Gamma&=& -\bigg[2 \zeta  \bigg(\eta_{0}^2 r^4-\eta_{0} r^2
\bigg(\eta_{2} r^2 \bigg(4 r^2 e^{2 \eta_{2} r^2}-5\bigg)+ \bigg(2
r^2 e^{2 \eta_{2} r^2}+\big(1-2
r^2\big)\\\nonumber&\times&e^{\eta_{2} r^2}-5\bigg)\bigg)+\eta_{2}^2
\bigg(2 r^4-4 r^6 e^{2 \eta_{2} r^2}\bigg)-\eta_{2} r^2\bigg(2 r^2
e^{2 \eta_{2} r^2}+\big(1-2 r^2\big)\\\nonumber&\times& e^{\eta_{2}
r^2}-2\bigg)-3\bigg(e^{\eta_{2}
r^2}-1\bigg)\bigg)\bigg]\bigg[-\big(3 \eta_{0} \eta_{2}
r^4+\eta_{2}^2 r^4+\eta_{2} r^2\big)3 \zeta -3\\\nonumber&\times&
e^{\eta_{2} r^2} \bigg(\zeta\big(2 \eta_{0} r^4+2 \eta_{2}
r^4\big)+q_{0}^2 r^4 \bigg)+6 \zeta r^4 (\eta_{0}+\eta_{2}) e^{2
\eta_{2} r^2} \bigg(\eta_{2} r^2\bigg)\bigg]^{-1},
\\\nonumber \Gamma_{r}&=& \bigg[\zeta  \bigg(\eta_{0} r^2
\bigg(2 r^2 e^{2 \eta_{2} r^2}-\big(2 r^2+1\big) e^{\eta_{2}
r^2}+3\bigg)+\eta_{2} r^2 \bigg(2 r^2 e^{2 \eta_{2} r^2}-e^{\eta_{2}
r^2}\\\nonumber&\times& \big(2 r^2+1\big)-1\bigg)-2
\bigg(e^{\eta_{2} r^2}-1\bigg)\bigg) \bigg(\zeta \big(2 \eta_{2}^2
r^4-\eta_{2} r^2\big)+e^{\eta_{2} r^2}\big(\zeta +q_{0}^2
\\\nonumber&\times&
r^4\big)\bigg)\bigg]\bigg[\bigg(-\zeta \big(3\eta_{0} \eta_{2}
r^4+\eta_{2}^2 r^4+\eta_{2} r^2\big)- e^{\eta_{2} r^2}\big(\zeta
\big(2 \eta_{0} r^4+2 \eta_{2}
r^4\big)+r^4\\\nonumber&\times&q_{0}^2 \big) +2 \zeta r^4
(\eta_{0}+\eta_{2}) e^{2 \eta_{2} r^2} \big(\eta_{2}
r^2\big)\bigg)\bigg(\zeta \big(3 \eta_{0} r^2+\eta_{2}
r^2\big)-e^{\eta_{2} r^2}\bigg(\zeta \big(2 \eta_{0}
r^4\\\nonumber&+& \eta_{0} r^2+2 \eta_{2} r^4+\eta_{2}
r^2\big)+q_{0}^2 r^4\bigg)+2 \zeta r^4 (\eta_{0}+\eta_{2}) e^{2
\eta_{2} r^2}\bigg)\bigg]^{-1},
\\\nonumber \Gamma_{t}&=&-\bigg[ \bigg(\zeta  \big(2 \eta_{2}^2 r^4-\eta_{2}
r^2 \big)+ e^{\eta_{2} r^2} \bigg(\zeta +q_{0}^2 r^4\bigg)\bigg)
\bigg(\zeta  \bigg(\eta_{0} \eta_{2} r^4-\eta_{0}^2 r^4-2 \eta_{0}
r^2\\\nonumber&-&\eta_{2} r^2\bigg)+ e^{\eta_{2} r^2} \big(2 q_{0}^2
r^4-\zeta \big)\bigg)\bigg]\bigg[r^6
\bigg(\zeta\eta_{2}+\zeta\big(\eta_{0} \eta_{2} r^2-\eta_{0}^2 r^2-2
\eta_{0}\big)+q_{0}^2
\\\nonumber&\times& r^2  e^{\eta_{2} r^2}\bigg) \bigg(\zeta
\bigg(\eta_{0}^2 \big(-\eta_{2} r^2\big)+\eta_{2}^2+\eta_{0}
\eta_{2} \bigg(\eta_{2} r^2-3 \bigg)\bigg)-q_{0}^2 R^6 e^{\eta_{2}
r^2}\bigg)\bigg]^{-1}.
\end{eqnarray}
\begin{figure}\center
\epsfig{file=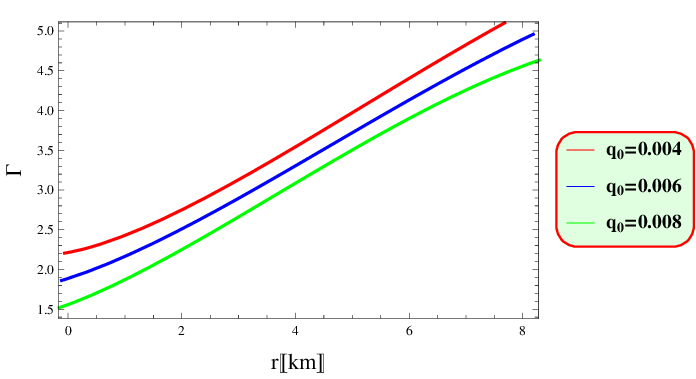,width=.5\linewidth}\epsfig{file=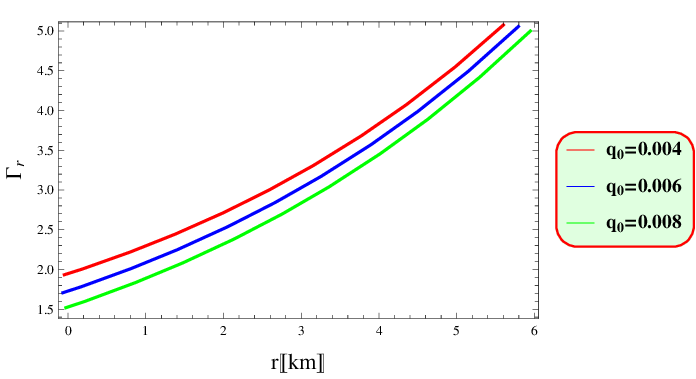,width=.5\linewidth}
\epsfig{file=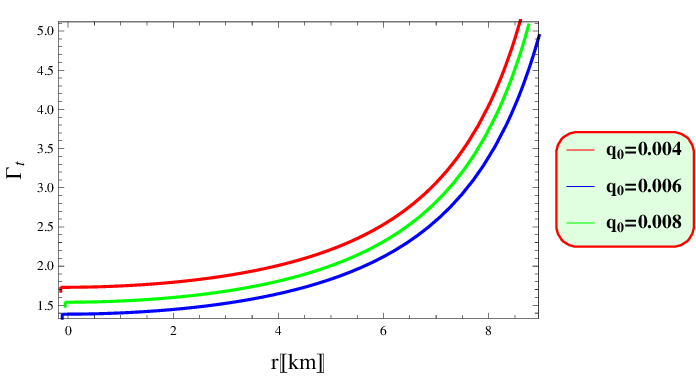,width=.5\linewidth}\caption{Graphical
representations of adiabatic index relative to $r$.}
\end{figure}

Figure \textbf{7} demonstrates that our $f(Q)$ gravity model
satisfies $\Gamma > \frac{4}{3}$, ensuring a stable anisotropic
model for the PS with various values of $q_{0}$. The TOV equation is
essential in theoretical astrophysics, particularly for
understanding the equilibrium structure of spherical stellar
objects. It describes the balance between gravitational forces and
internal pressure, which is crucial for maintaining the stability of
a star under its own gravity. The TOV equation is expressed as
\cite{50}
\begin{equation}\label{40}
M_{G}(r) e^{\frac{\psi-\xi}{2}} \frac{1}{r^2} (\rho + p_r) +
\frac{dp_r}{dr}-\frac{2}{r} (p_t - p_r) = 0,
\end{equation}
here the gravitational mass is defined as \cite{51}
\begin{equation}\nonumber
M_{G}(r) = 4\pi \int (\mathbb{T}^t_t- \mathbb{T}^r_r -
\mathbb{T}^\phi_\phi- \mathbb{T}^\theta_\theta) r^2
e^{\frac{\psi+\xi}{2}} dr.
\end{equation}
Substituting these values and integrating this equation, we get
\begin{equation}\nonumber
M_{G}(r) = \frac{1}{2} r^2 e^{\frac{\xi-\psi}{2}} \psi^{\prime}.
\end{equation}
Inserting this value into Eq.\eqref{40}, we find that
\begin{equation}\nonumber
\frac{1}{2} \psi^{\prime} (\rho + p_r) + \frac{dp_r}{dr} -
\frac{2}{r} (p_t - p_r) = 0.
\end{equation}
We examine the hydrodynamic equilibrium of our model by applying the
TOV equation within the framework of $f(Q)$ gravity
\begin{equation}\nonumber
F_{a}+F_{g}+F_{h}+F_{Q}= 0.
\end{equation}
Analyzing the stability of a model under various forces is crucial.
For an anisotropic charged compact object, stability is assessed by
examining four key force components. The gravitational force
$F_{g}$, the hydrostatic force $F_{h}$, the anisotropic force
$F_{a}$ and the electromagnetic force $F_{Q}$. To ensure the model's
stable equilibrium, these forces must be properly balanced. In this
study, we use the generalized TOV equation to analyze stability.
These can be defined as
\begin{eqnarray}\nonumber
F_{a}&=&2\frac{\chi}{r},\quad F_{g}= \frac{\psi^{\prime} (p_{r}+\rho
)}{2},\quad F_{h}=-p_{r}^{\prime},\\\nonumber
F_{Q}&=&p_{r}^{\prime}+\frac{\xi^{\prime}}{2}(\rho-p_{r})-\frac{2}{r}(p_{t}-p_{r}).
\end{eqnarray}
By substituting the values, we get
\begin{eqnarray}\nonumber
F_{a}&=& \frac{1}{r^3}\bigg[2 e^{-\eta_{2} r^2} \bigg(\zeta
\big(\eta_{0}^2 r^4-\eta_{0} \eta_{2} r^4+5 \eta_{0} r^2 \big)-
e^{\eta_{2} r^2} \bigg( \big(2 \eta_{0} r^4+\eta_{0} r^2+2 \eta_{2}
r^4\\\nonumber&+&\eta_{2} r^2\big)\zeta+2 q_{0}^2 r^4 \bigg)+2 \zeta
r^4 (\eta_{0}+\eta_{2}) e^{2 \eta_{2} r^2}\bigg)\bigg],
\\\nonumber
F_{g}&=& -\frac{1}{2 r}\bigg[\eta_{0} \zeta  e^{-\eta_{2} r^2}
\bigg(\eta_{0} r^2 \bigg(2 r^2 e^{2 \eta_{2} r^2}-\big(2 r^2+1\big)
e^{\eta_{2} r^2}+3\bigg)+\eta_{2} r^2\bigg( e^{2 \eta_{2} r^2}
\\\nonumber&\times&2 r^2-\big(2
r^2+1\big) e^{\eta_{2} r^2}-1\bigg)-2 R^2 \bigg(e^{\eta_{2}
r^2}-1\bigg)\bigg)\bigg],
\\\nonumber
F_{h}&=& 2 r \bigg(q_{0}^2-\big(\zeta e^{-\eta_{2} r^2}
\big(\eta_{0}^2 \big(-\eta_{2} r^2\big)+\eta_{0} \eta_{2}
\big(\eta_{2} r^2-3\big)+\eta_{2}^2\big)\big)\bigg),
\\\nonumber
F_{Q}&=&\frac{1}{2 r^3}\bigg[e^{-\eta_{2} r^2} \bigg(4 \eta_{2}r^2
\bigg(2 e^{2 \eta_{2} r^2} (\eta_{0}+\eta_{2}) \zeta
r^4+\big((3\eta_{0}+\eta_{2}) r^2\big) \zeta-e^{\eta_{2} r^2}
\\\nonumber&\times&\bigg(q_{0}^2 r^4+\big((\eta_{0}+\eta_{2}) \big(2 r^2+1\big) r^2\big) \zeta
\bigg)\bigg) +4 \bigg(-4 e^{2 \eta_{2}
r^2}\zeta(\eta_{0}+\eta_{2})r^2\\\nonumber&\times&
 \big(\eta_{2} r^2\big) -(3\eta_{0}+\eta_{2})
\zeta + \bigg(\eta_{2}^2 \big(2 r^2+1\big) \zeta r^2+\big(2 q_{0}^2
r^2+\eta_{0} \big(4 r^2+1\big) \zeta \big)\\\nonumber&+&
\big(q_{0}^2  r^4+\big(2 r^2+1\big) \big(\eta_{0} r^2+2 \big)\zeta
\big)\eta_{2} \bigg)e^{\eta_{2} r^2}\bigg) r^2-4 \bigg(2 e^{2
\eta_{2} r^2} (\eta_{0}+\eta_{2}) \zeta r^4\\\nonumber&+&
\big(\eta_{0}^2 r^4-\eta_{0} \eta_{2} r^4+5 \eta_{0} r^2\big)\zeta
-e^{\eta_{2} r^2}\big(2 q_{0}^2 r^4+\big(2 \eta_{0} r^4+2 \eta_{2}
r^4+\eta_{0} r^2
\\\nonumber&+&\eta_{2} r^2\big) \zeta \big)\bigg)(\eta_{0}+\eta_{2}) \zeta r^4-\big((3\eta_{0}+\eta_{2})
r^2\big) \zeta + \big(\big((\eta_{0}+\eta_{2}) \big(2 r^2+1\big)
r^2\\\nonumber&+&q_{0}^2 r^4\big) \zeta \big)e^{\eta_{2}
r^2}\bigg)-\bigg[r  \bigg(\eta_{2} \bigg(2 e^{2 \eta_{2} r^2}
r^2-e^{\eta_{2} r^2} \big(2 r^2+1\big)-1\bigg) r^2+\eta_{0}\bigg(2
\\\nonumber&\times&
e^{2 \eta_{2} r^2} r^2-e^{\eta_{2} r^2} \big(2 r^2+1\big)+3\bigg)
r^2-2 \bigg(e^{\eta_{2} r^2}-1\bigg) \bigg) \zeta \bigg(2\bigg(2
e^{2 \eta_{2} r^2} \zeta r^4\\\nonumber&\times&r(\eta_{0}+\eta_{2})
\big(\eta_{2} r^2\big) -\big(\eta_{2}^2 r^4+3\eta_{0} \eta_{2}
r^4+\eta_{2} r^2\big) \zeta- e^{\eta_{2} r^2} \bigg(q_{0}^2
r^4+\big(2\eta_{0} r^4
\\\nonumber&+&2 \eta_{2} r^4\big) \zeta
\bigg)\bigg)\bigg(e^{\eta_{2} r^2}\bigg(2 q_{0}^2
r^2+\eta_{2}\big(q_{0}^2 r^4+\zeta \big)\bigg) +\eta_{2} \big(4
\eta_{2} r^2\big) \zeta-\bigg(2 r \bigg(e^{\eta_{2}
r^2}\\\nonumber&\times& \big(q_{0}^2 r^4+\zeta \big) \bigg)+\big(2
\eta_{2}^2 r^4-\eta_{2} r^2\big) \zeta \bigg) \bigg(2 e^{2 \eta_{2}
r^2} \big(2 \eta_{2}^2 r^4+5 \eta_{2}r^2+2\big)\zeta r^2
(\eta_{0}\\\nonumber&+&\eta_{2})-\eta_{2}\big(6 \eta_{0} r^2+2
\eta_{2} r^2\big) \zeta-e^{\eta_{2} r^2}\bigg(2 \eta_{2}^2 \zeta
r^4+2\big(q_{0}^2+2 \eta_{0}\zeta \big)r^2+\eta_{2}
\\\nonumber&\times&  \big(q_{0}^2 r^4+\big(\eta_{0} r^4+\big(4
r^2-1\big)
\big)\zeta\big)\bigg)\bigg)\bigg)\bigg)\bigg]\bigg[\bigg(\zeta\big(\eta_{2}^2
r^4+3\eta_{0} \eta_{2} r^4+\eta_{2}
 r^2\big)\\\nonumber&+& 2 e^{2 \eta_{2} r^2}
(\eta_{0}+\eta_{2})\big(\eta_{2} r^2\big)  \zeta r^4 +e^{\eta_{2}
r^2}\bigg(q_{0}^2 r^4+\big(2\eta_{0} r^4+2 \eta_{2} r^4\big) \zeta
\bigg)\bigg)^2\bigg]^{-1}\bigg)\bigg].
\end{eqnarray}
We have illustrated the condition for stable equilibrium using a
graphical representation. Figure \textbf{8} shows that equilibrium
is attained when the combined forces of $F_{a}$, $F_{g}$, $F_{h}$
and $F_{Q}$ sum to zero. This confirms that the PS model remains
stable across different values of $q_{0}$. The figure presents two
distinct graphs that show the behavior of $F_{Q}$. In one where
$F_{Q} = 0$, it indicates no electromagnetic force and the stability
of the system depends solely on the balance of the gravitational,
hydrostatic and anisotropic forces. If these forces are in balance,
the system can remain stable, though the absence of electromagnetic
force might lead to variations in pressure and structure. The other
graph shows $F_{Q} \neq 0$, where the electromagnetic force provides
additional repulsion, helps to counteract gravitational collapse and
plays a key role in maintaining stability.
\begin{figure}\center
\epsfig{file=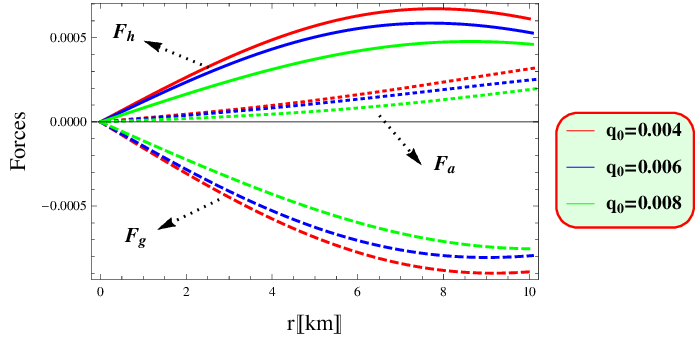,width=.5\linewidth}\epsfig{file=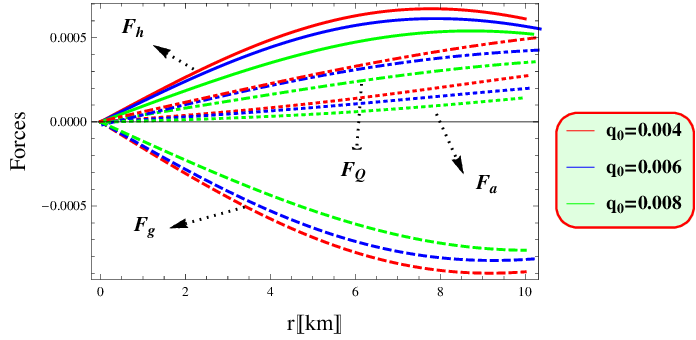,width=.5\linewidth}
\caption{Graphs of TOV equation as functions $r$.}
\end{figure}

\section{Equation of State Parameter and Compactness}

Here, we apply the subsequent equations \cite{36}
\begin{equation}\nonumber
p_r(\rho) \approx v^2_r (\rho - \rho_I), \quad p_t(\rho) \approx
v^2_t (\rho - \rho_{II}).
\end{equation}
In this model, $\rho_I$ and $\rho_{II}$ represent the densities at
the surface of the star, related to the radial and tangential
pressures, respectively. While $\rho_I$ can make $p_r$ zero,
$\rho_{II}$ does not have the same effect on the $p_t$. The model
helps us figure out both the speed of sound and the surface density
of the star. For instance, with $q_{0} = 0.004$, inserting
Eqs.\eqref{38} and \eqref{39}, gives us $v_r^2 \approx 0.793,~v_t^2
\approx 0.851$, $ \rho_I \approx 5.01 \times 10^{14}$ g/cm$^{3} $
and $ \rho_{II} \approx 3.99 \times 10^{14}$ g/cm$^{3} $. Similarly,
for $q_{0} = 0.006$, the values are $ v_r^2 \approx 0.789,~v_t^2
\approx 0.842$, $ \rho_I \approx 5.04 \times 10^{14}$ g/cm$^{3} $
and $ \rho_{II} \approx 4.2 \times 10^{14}$ g/cm$^{3}$. Also, for
$q_{0} = 0.008$,  we find $ v_r^2 \approx 0.785,~v_t^2 \approx
0.717$, $ \rho_I \approx 0.833 \times 10^{14}$ g/cm$^{3}$ and
$\rho_{II} \approx 4.4 \times 10^{14}$ g/cm$^{3}$.
\begin{figure}\center
\epsfig{file=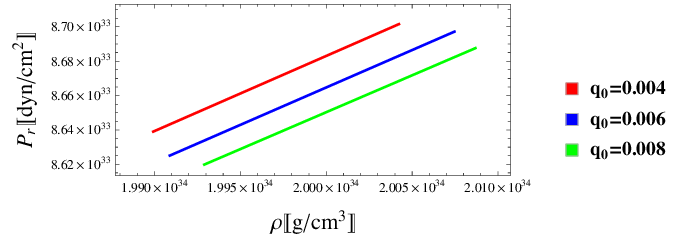,width=.46\linewidth}\epsfig{file=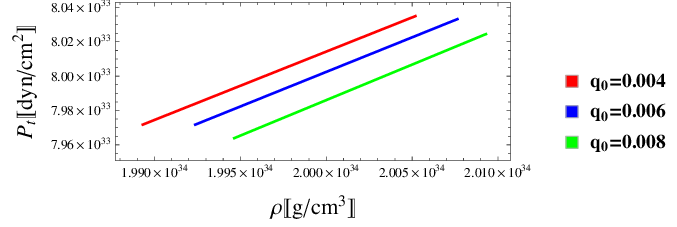,width=.48\linewidth}
\epsfig{file=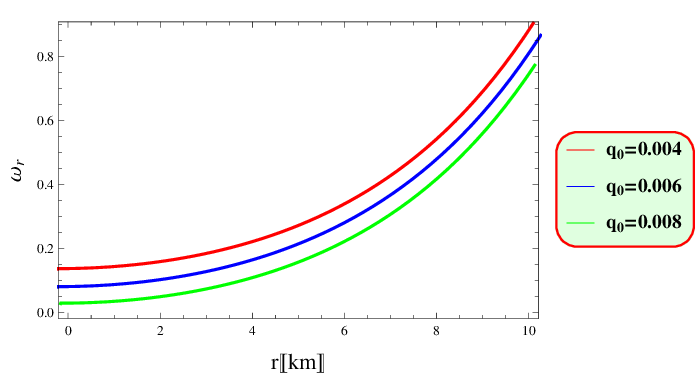,width=.5\linewidth}\epsfig{file=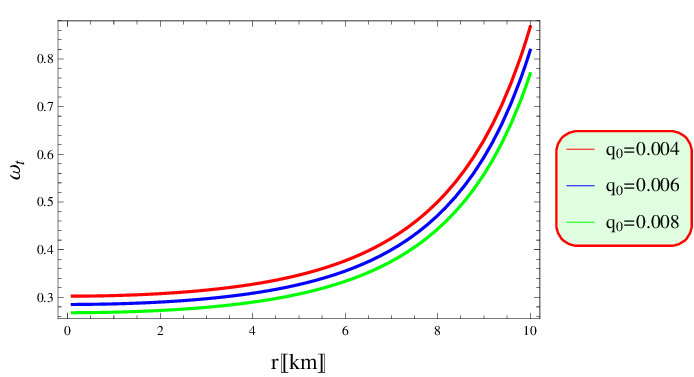,width=.5\linewidth}
\caption{Graphs of EoS versus $r$.}
\end{figure}

The EoS parameter, denoted as $\omega$, is expressed as $\omega =
\frac{p}{\rho}$. In astrophysics and cosmology, this parameter is
crucial for understanding the behavior of different components of
the universe. A valid model requires that both the radial EoS
parameter $(\omega_{r}=\frac{p_{r}}{\rho})$ and the tangential EoS
parameter $(\omega_{t}=\frac{p_{t}}{\rho})$ to be within the range
of $[0,1]$ \cite{46}. By substituting Eqs.\eqref{32}-\eqref{34} into
the above formulas, we obtain
\begin{eqnarray}\nonumber
\omega_{r}&=& \bigg[-e^{\eta_{2} r^2} \bigg(\zeta \big(2 \eta_{0}
r^4+ \eta_{0} r^2+2 \eta_{2} r^4+\eta_{2} r^2\big)+q_{0}^2
r^4\bigg)+2 \zeta  r^4 e^{2 \eta_{2} r^2} \\\nonumber &\times&
(\eta_{0}+\eta_{2}) +\zeta\big(3\eta_{0} r^2+\eta_{2}
r^2\big)\bigg]\bigg[ e^{\eta_{2} r^2} \bigg(q_{0}^2 r^4-\zeta
\bigg)-\zeta \big(2 \eta_{2} r^2\big)\bigg]^{-1},
\\\nonumber
\omega_{t}&=& \bigg[r^2 \bigg(\bigg(\zeta e^{-\eta_{2} r^2}
\bigg(\eta_{0} r^2 (\eta_{0}-\eta_{2})+
(2\eta_{0}-\eta_{2})\bigg)\bigg)-q_{0}^2 r^2\bigg)\bigg]\\\nonumber
&\times&\bigg[\zeta e^{-\eta_{2} r^2} \bigg(2 \eta_{2}
r^2-1\bigg)+\zeta -q_{0}^2 r^4\bigg]^{-1}.
\end{eqnarray}
Figure \textbf{9} displays the best-fit EoS for the PS. It shows how
density and radial pressure change with different values of $q_{0}$,
align well with a linear EoS pattern. Similarly, the tangential EoS
exhibits a strong relationship with a linear model. These EoSs are
primarily accurate near the star's center and throughout its
interior, within the the specified range $[0,1]$ \cite{47}.
\begin{figure}\center
\epsfig{file=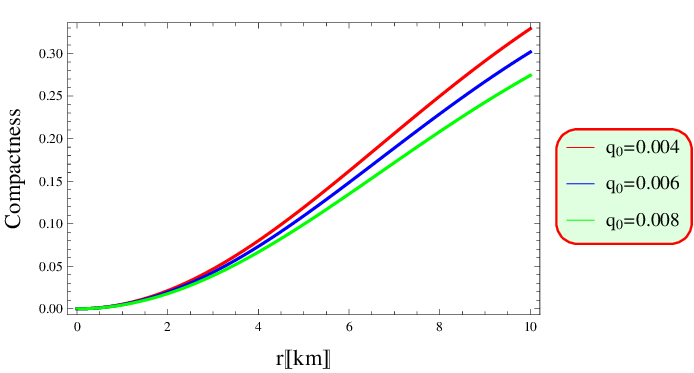,width=.5\linewidth} \caption{Graph of
compactness versus $r$.}
\end{figure}

The compactness function $\varpi = 2M(r)/r $ is essential for
evaluating the star's stability. For a star to be stable, its
compactness must be less than $4/9$ \cite{40}. This criterion helps
to make sure that the star's gravity is not strong enough to cause
it to collapse. Therefore, it is important to check that $\varpi$
remains below this limit to ensure the star's stability.
\begin{equation}\nonumber
\varpi= 4 \pi  \zeta  \bigg(1-e^{-\eta_{2} r^2}\bigg)-\frac{4}{5}
\pi  q_{0}^2 r^4.
\end{equation}
Figure \textbf{10} demonstrates that the compactness increases
gradually while remaining within the specified limit.

\section{Summary and Discussion}

In this paper, we have examined how the $f(Q)=\zeta Q$ gravity model
affects the structure and stability of charged PS. By considering an
anisotropic fluid and using the KB ansatz for the star's interior,
we have determined key constraints on the model parameters,
especially the value of $\zeta$. We have investigated the charge
intensity $q_0$ which influenced our model through both numerical
calculations and graphical representations. The key features are
outlined as follows.
\begin{itemize}
\item
We have found that the density and both radial/tangential pressures
are highest at the core of the star and decrease towards the
surface. Importantly, the radial pressure drops to zero at the
surface of the star. As the charge intensity increases, the energy
density and pressure go down. However, the steadily decreasing
pattern of these profiles remains consistent in this model (Figure
\textbf{1}).
\item
The anisotropy consistently rises towards the star's surface while
still fulfilling stability requirements (Figure \textbf{2}). This
behavior aligns with theoretical predictions and confirms the
physical viability of the proposed stellar model.
\item
The mass of the PS increases consistently and uniformly as its
radius expands (Figure \textbf{3}). This behavior aligns with the
stability of the PS.
\item
We have observed that the redshift function keeps increasing for
different values of $q_{0}$ (Figure \textbf{4}). It always stays
below the maximum value of 5.211 which is the highest value obtained
from observational data.
\item
We have examined the energy conditions for various values of $q_{0}$
and demonstrated that the PS model satisfies all the energy
requirements (Figure \textbf{5}). This confirms that our model is
physically viable.
\item
It is found for various values of $q_{0}$ the expressions $(0\leq
v_r^2 \leq 1)$ and $(0 \leq v_t^2 \leq 1)$ are satisfied ((Figure
\textbf{6})) that confirm the stability and causality requirements.
Furthermore, our analysis reveals that the condition $(0\leq
v_t^2-v_r^2\leq 1)$ is satisfied throughout the pulsar's interior
which indicates a stable anisotropic stellar structure.
\item
We have observed that the adiabatic index $\Gamma$ is greater than
$\frac{4}{3}$, indicating that the PS remains physically stable for
different values of $q_{0}$ (Figure \textbf{7}).
\item
The equilibrium is achieved when the total of the forces $F_{a}$,
$F_{g}$, $F_{h}$ and $F_{Q}$ equals to zero (Figure \textbf{8}).
This ensures a stable model for the PS for various values of
$q_{0}$.
\item
The EoS for the PS shows a strong linear relationship between
density and radial pressure across all values of $q_{0}$. The
tangential pressure also fits well with a linear model. These
relationships hold true near the center of the star and throughout
its interior, especially within the range of 0 to 1 (Figure
\textbf{9}).
\item
The compactness gradually rises inside the star and remains within
the limit of $\varpi <4/9$ for various values of $q_{0}$ (Figure
\textbf{10}).
\end{itemize}
It is noteworthy that $f(Q)$ gravity enables the construction of
realistic models adhering to essential principles for static
spherically symmetric spacetime. We have successfully developed a
stable and physically viable model of a charged PS using $f(Q)$
gravity. Our findings show that this theory meets theoretical
stability criteria and aligns well with observational data. The
inclusion of charge significantly influences the star's structure,
affecting its mass, radius and stability. Despite these variations,
the charged model remains stable and physically viable. Our results
are consistent with recent work in $f(Q)$ gravity \cite{4aa},
reinforcing the validity of pulsar studies within this framework and
suggesting future research could provide new insights into gravity
and compact objects.\\\\
\textbf{Data Availability:} No data was used for the research
described in this paper.

\end{document}